\documentclass[aps,prl,twocolumn,groupedaddress,nofootinbib,showpacs]{revtex4}
\usepackage[dvips]{graphicx}
\newcommand{\BlackHat}{{\sc BlackHat}}
\newcommand{\SHERPA}{{\sc SHERPA}}

\newcommand{\AMEGIC}{{\sc AMEGIC++}}

\newif\ifdraft
\drafttrue
\newif\ifpreprint
\preprinttrue

\def\fig#1{fig.~{\ref{#1}}}
\def\Fig#1{Fig.~{\ref{#1}}}

\def\tab#1{table~{\ref{#1}}}

\def\bom#1{{\mbox{\boldmath $#1$}}}
\def\Wz{$W\,\!+\,0$}
\def\Wj{$W\,\!+\,1$}

\def\Wjjj{$W\,\!+\,3$}
\def\Wjjjj{$W\,\!+\,4$}
\def\Wmjjjj{$W^-\,\!+\,4$}
\def\Wpjjjj{$W^+\,\!+\,4$}

\def\Wjjjx{$W\,\!+\,1,2,3$}
\def\Wjn{$W\,\!+\,n$}
\def\Wmjn{$W^-\,\!+\,n$}
\def\Wmjnm{$W^-\,\!+\,(n\!-\!1)$}

\def\Zjjjj{$Z\,\!+\,4$}

\def\Zgamjjj{$Z,\gamma^*\,\!+\,3$}

\def\Zjjjj{$Z\,\!+\,4$}

\def\jet{{\rm jet}}
\def\pt{p_T}

\def\ETsl{{\s E}_T}

\def\HTpartonicp{{\hat H}_T'}

\newbox\charbox
\newbox\slabox
\def\s#1{{      
        \setbox\charbox=\hbox{$#1$}
        \setbox\slabox=\hbox{$/$}
        \dimen\charbox=\ht\slabox
        \advance\dimen\charbox by -\dp\slabox
        \advance\dimen\charbox by -\ht\charbox
        \advance\dimen\charbox by \dp\charbox
        \divide\dimen\charbox by 2
        \raise-\dimen\charbox\hbox to \wd\charbox{\hss/\hss}
        \llap{$#1$}
}}

\begin{document}

\title{
\ifpreprint
\hbox{\rm\small
SB/F/385-10$\null\hskip 2.3cm \null$
UCLA/10/TEP/106$\null\hskip 2.3cm \null$
SLAC--PUB--14222$\null\hskip 2.3 cm \null$
MIT CTP-4169\break}
\hbox{\rm\small IPPP/10/68 $\null\hskip 2.4cm \null$
Saclay IPhT--T10/111$\hskip 2.cm \null$
NIKHEF-2010-022$\hskip 1.7cm \null$
CERN-TH/2010-184\break}
\hbox{$\null$\break}
\fi
Precise Predictions for $\bom{W}$\,+\,4 Jet Production
at the Large Hadron Collider
}

\author{C.~F.~Berger${}^{a}$,\, Z.~Bern${}^b$,\,
L.~J.~Dixon${}^{c,d}$,\, F.~Febres Cordero${}^e$,\, D.~Forde${}^{c,f}$,\,
T.~Gleisberg${}^d$,\,  H. Ita${}^b$,\,
D.~A.~Kosower${}^{g}$\,
and D.~Ma\^{\i}tre${}^{h}$ 
\\
$\null$
\\
${}^a$Center for Theoretical Physics, MIT,
      Cambridge, MA 02139, USA \\
${}^b$Department of Physics and Astronomy, UCLA, Los Angeles, CA
90095-1547, USA \\
${}^c$Theory Division, Physics Department, CERN, CH--1211 Geneva 23, 
    Switzerland\\
${}^d$SLAC National Accelerator Laboratory, Stanford University,
             Stanford, CA 94309, USA \\
${}^e$Departamento de F\'{\i}sica, Universidad Sim\'on Bol\'{\i}var, 
 Caracas 1080A, Venezuela\\
${}^f$NIKHEF Theory Group, Science Park 105, NL--1098~XG
  Amsterdam, The Netherlands\\
${}^g$Institut de Physique Th\'eorique, CEA--Saclay,
          F--91191 Gif-sur-Yvette cedex, France\\
${}^h$Department of Physics, University of Durham,
          Durham DH1 3LE, UK\\
}

\begin{abstract}
We present the next-to-leading order (NLO) QCD results for \Wjjjj-jet
production at hadron colliders.
This is the first hadron collider process with five final-state objects
to be computed at NLO. It represents an important background to
many searches for new physics at the energy frontier.
Total cross sections, as well as
distributions in the jet transverse momenta\ifpreprint{}
and in the total transverse energy $H_T$\fi,
are provided for the initial LHC energy of $\sqrt{s} = 7$ TeV.  We use a
leading-color approximation, known to be accurate to 3\% for $W$
production with fewer jets.  The calculation uses the \BlackHat{} library
along with the \SHERPA{} package.
\end{abstract}

\pacs{12.38.-t, 12.38.Bx, 13.87.-a, 14.70.Fm \hspace{1cm}}

\maketitle


The first data and analyses emerging from experiments at
the Large Hadron Collider (LHC) emphasize the need for
reliable theoretical calculations in searches
for new physics beyond the Standard Model.  In many channels, 
new-physics signals can hide in broad distributions underneath
Standard Model backgrounds.  Extraction of a signal
will require accurate
predictions for the background
processes, for which next-to-leading order (NLO) cross sections
in perturbative QCD are crucial.
The past few years have seen rapid progress in NLO QCD for the LHC.
Several important processes involving four 
final-state objects (including jets) have been
computed~\cite{PRLW3BH,EMZW3Tev,W3jDistributions,MZ3j,TeVZ,OtherNLO}.

In this Letter, we present results for the first of a new class
of processes, involving five final-state objects: inclusive \Wjjjj-jet
production, using a leading-color approximation for the virtual
terms that has been validated for processes with fewer jets.
This process has been studied since the
early days of the Tevatron, where it was the dominant
background to top-quark pair production.  At the LHC, it will be an
important background to many new physics searches 
involving missing energy,
as well as to precise top-quark measurements.

In previous papers~\cite{W3jDistributions,TeVZ} we presented the
first complete results for hadron-collider production of a $W$ or
$Z$ boson in association with three jets at NLO in the strong coupling
$\alpha_s$.  (Other NLO results for \Wjjj{} jets have used
various leading-color approximations~\cite{PRLW3BH,EMZW3Tev,MZ3j}.)
We performed detailed comparisons to Tevatron data~\cite{TevatronPapers}.
The sensitivity to the unphysical scale used to define
$\alpha_s$ and the parton distributions is reduced
from around 40\% at leading order (LO) 
to 10$\sim$20\% at NLO, and the NLO results agree well with the data.
At the LHC, a much wider range of kinematics will be probed,
making NLO studies even more important.


\begin{figure}[t]
\includegraphics[clip,scale=0.4]{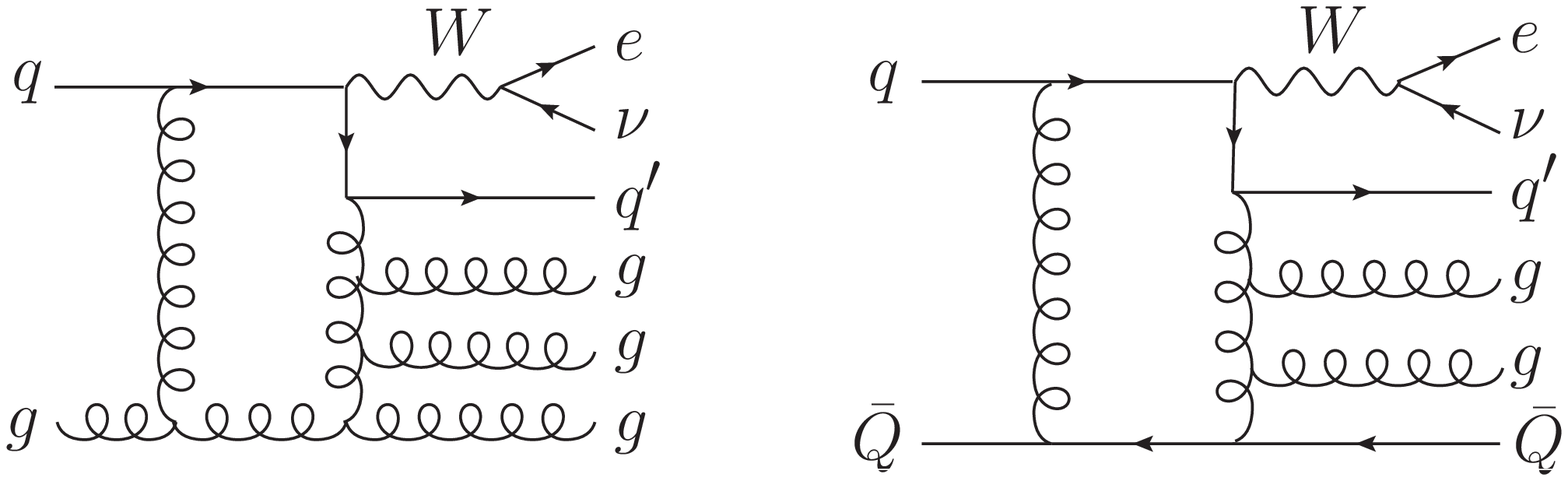}
\caption{Sample diagrams for the seven-point loop amplitudes for
$q g \rightarrow W \, q' \! g g g$ and 
$q \bar Q \rightarrow W \, q' \! g g \bar Q$, followed by $W \to e\nu$.}
\label{lcdiagramsFigure}
\end{figure}

The computation of hadron collider processes with complex final states
at NLO has long been a challenge to theorists.  The evaluation of the
one-loop (virtual) corrections has been a longstanding bottleneck.
Feynman-diagram techniques suffer from rapid growth in complexity as
the number of legs increases.  On-shell
methods~\cite{UnitarityMethod,BCFUnitarity,BCFW,Bootstrap,OPPEtc,OnShellReview},
in contrast, rely on the unitarity and factorization properties of
scattering amplitudes to generate new amplitudes from
previously-computed ones.  Such methods scale very well as the number
of external legs increases, offering a solution to these difficulties.

We use the same basic setup as in our earlier
computations~\cite{W3jDistributions,TeVZ} of \Wjjj-jet 
and \Zgamjjj-jet production. The virtual contributions are computed
using on-shell methods via the \BlackHat{} package~\cite{BlackHatI}. 
We show representative virtual diagrams in \fig{lcdiagramsFigure}. 
We use a leading-color approximation in the finite virtual
contributions\footnote{Our definition of leading-color terms follows 
ref.~\cite{TeVZ}; it includes virtual quark loops in addition 
to the terms identified in ref.~\cite{PRLW3BH}.}, while
keeping the full color dependence in all other contributions.
We have confirmed that this approximation is an excellent one for
\Wjjjx-jet production, shifting the total cross section
by about 3\%, which is significantly smaller than uncertainties
from parton distributions or higher-order terms in $\alpha_s$.
Subleading-color corrections to \Wjjjj-jet production should also be
small.  We include the full $W$ Breit-Wigner resonance; decays to leptons
retain all spin correlations.

The remaining NLO ingredients, the real-emission and dipole-subtraction
terms~\cite{CS}, are computed by \AMEGIC{}~\cite{Amegic}, part of
the \SHERPA{} package~\cite{Sherpa}.  We also use \SHERPA{} to perform
phase-space integration.  The efficiency of the integrator 
has been improved significantly with respect to 
ref.~\cite{W3jDistributions} through the use of QCD antenna
structures~\cite{AntennaIntegrator,GleisbergIntegrator}.
\BlackHat{} computes the
real-emission
tree amplitudes using on-shell recursion
relations~\cite{BCFW}, along with efficient
analytic forms extracted from ${\cal N}=4$ super-Yang-Mills
theory~\cite{DrummondTrees}.


Compared to LO, NLO cross sections and distributions generally
depend much less on the common (unphysical)
renormalization and factorization scale $\mu$.
However, it is still important to select a scale characteristic of
the typical kinematics.
A scale that performs well for many distributions is the
total partonic transverse energy. We set
$\mu = \HTpartonicp/2$, where 
$\HTpartonicp = \sum_j p_T^j + E_T^W$;
the sum runs over all final-state partons $j$, and
$E_T^W = \sqrt{M_W^2+(p_T^W)^2}$ is the transverse energy of the $W$
boson\footnote{In refs.~\cite{W3jDistributions,TeVZ} we used the
scalar sum of the decay leptons' transverse energies instead of $E_T^W$.
The present choice is preferred for studies of $W$
polarization effects~\cite{W3jDistributions,Polarization}.}. 
Refs.~\cite{Bauer,MZ3j} present other satisfactory choices.
We follow the conventional procedure
of varying the chosen central scale up and down by a factor of two
to construct scale-dependence bands, taking
the minimum and maximum of the observable evaluated at five
values: $\mu/2, \mu/\sqrt2, \mu, \sqrt2\mu, 2\mu$.

Fixed-order perturbation theory may break down in special
kinematic regions, where large logarithms of scale ratios emerge.
For instance, threshold logarithms can affect production at
very large mass scales, which can be reached in inclusive single-jet
production~\cite{deFV}.  Using this study one can argue~\cite{TeVZ}
that at the mass scales probed in \Wjjjj-jet production,
such logarithms should remain quite modest.
Similarly, the sort of large logarithms arising in
vector-boson production in association with a single jet~\cite{RSS}
do not appear in the case of multiple jets.  Tighter
cuts may isolate regions which would require a reassessment
of potentially-large logarithms.

\begin{figure*}[t]
\includegraphics[clip,scale=0.5]{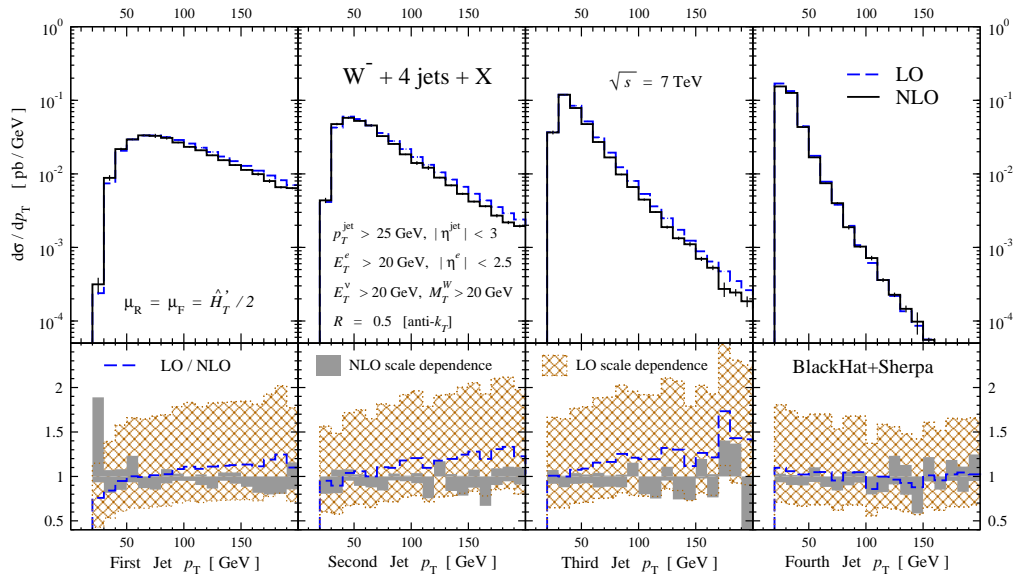}
\caption{A comparison of the $p_T$ distributions of the leading four
jets in \Wmjjjj{}-jet production at the LHC at $\sqrt{s}=7$~TeV.  In
the upper panels the NLO distribution is the solid (black) histogram
and the LO predictions are shown as dashed (blue) lines.  The thin
vertical line in the center of each bin (where visible) gives its
numerical (Monte Carlo) integration error.  The lower panels show the
LO distribution and LO and NLO scale-dependence bands 
normalized to the central NLO prediction.
The bands are shaded (gray) for NLO and cross-hatched
(brown) for LO. }
\label{W4ptFigure}
\end{figure*}

\begin{table*}
\vskip .4 cm
\begin{tabular}{||c||c|c||c|c||c|c||}
\hline
no. jets &  $W^-$ LO  & $W^-$ NLO & $W^+/W^-$ LO &  $W^+/W^-$ NLO & $W^- n/(n\!-\!1)$ LO & $W^- n/(n\!-\!1)$ NLO  \\
\hline
0 & $1614.0(0.5)^{+208.5}_{-235.2}$  & $2077(2)^{+40}_{-31}$ &   $1.656(0.001)$  &  $1.580(0.004)$ & ---  & ---  
 \\
\hline
1 & $\;  264.4(0.2)^{+22.6}_{-21.4} \;$ &   $ 331(1)^{+15}_{-12}\; $ &$1.507(0.002)$  & $1.498(0.009)$ & $0.1638(0.0001)^{+0.044}_{-0.031}$ & $0.159(0.001)$  
 \\
\hline
2 & $\;73.14(0.09)^{+20.81}_{-14.92} \;$ & $78.1(0.5)^{+1.5}_{-4.1} $ & $1.596(0.003)$  & $1.57(0.02)$ & $0.2766(0.0004)^{+0.051}_{-0.037}$ & $0.236(0.002)$
 \\
\hline
3 & $\;17.22(0.03)^{+8.07}_{-4.95}\;$  &  $\; 16.9(0.1)^{+0.2}_{-1.3}\;$ & $1.694(0.005)$  &  $1.66(0.02)$ & $0.2354(0.0005)^{+0.034}_{-0.025}$ & $0.216(0.002)$
 \\
\hline
4 & $\; 3.81(0.01)^{+2.44}_{-1.34} \;$  &  $ \; 3.55(0.04)^{+0.08}_{-0.30}\; $  &  $1.812(0.001)$  & $1.73(0.03)$ &$0.2212(0.0004)^{+0.026}_{-0.020}$ & $0.210(0.003)$
 \\
\hline 
\end{tabular} 
\caption{Total cross sections in pb for \Wjn{} jet production at the
LHC at $\sqrt{s}=7$~TeV, using the  anti-$k_T$ jet algorithm with  $R=0.5$.
The NLO result for \Wjjjj{} jets uses the leading-color
approximation discussed in the text.  The fourth and fifth columns
give the cross-section ratios for $W^+$ production to $W^-$ production.
The last column gives the ratios of the cross section
for the given process to that with one jet less.
The numerical
  integration uncertainty is in parentheses, and the scale dependence
  is quoted in super- and subscripts.
\label{CrossSectionAnti-kt-R5Table} 
}
\end{table*}

\begin{table*}
\vskip .4 cm
\begin{tabular}{||c||c|c||c|c||c|c||}
\hline
no.\ jets &  $W^-$ LO  & $W^-$ NLO & $W^+/W^-$ LO &  $W^+/W^-$ NLO & $W^- n/(n\!-\!1)$ LO  & $W^- n/(n\!-\!1)$ NLO
 \\
\hline
0   & $1614.0(0.5)^{+208.5}_{-235.2}$  &  $2077(2)^{+40}_{-31}$ & $1.656(0.001)$ &    $1.580(0.004)$ & --- & --- 
 \\
\hline
1   & $\; 264.4(0.2)^{+22.6}_{-21.4} \;$ &  $ 324(1)^{+14}_{-11} $ &  $1.507(0.002)$  & $1.499(0.009)$ &$0.1638(0.0001)^{+0.044}_{-0.031}$ & $0.156(0.001)$

 \\
\hline
2  & $\;74.17(0.09)^{+21.08}_{-15.12} \;$  &  $76.2(0.5)^{+0.8}_{-3.4} $ &  $1.597(0.003)$   &   $1.56(0.02)$ & $0.2805(0.0004)^{+0.051}_{-0.038}$ & $0.235(0.002)$
 \\
\hline
3  & $\; 18.42(0.03)^{+8.61}_{-5.29} \;$  &  $17.0(0.1)^{+0.0}_{-1.0} $ &  $1.694(0.005)$&  $1.66(0.02)$ & $0.2483(0.0005)^{+0.036}_{-0.026}$ & $0.223(0.002)$
 \\
\hline
4  & $\; 4.41(0.01)^{+2.82}_{-1.55} \;$  &  $\;  3.81(0.04)^{+0.00}_{-0.44}\;$ &  $1.814(0.001)$  & $1.76(0.03)$ & $0.2394(0.0004)^{+0.028}_{-0.021}$ & $0.224(0.003)$
 \\
\hline 
\end{tabular} 
\caption{The same quantities as in~\tab{CrossSectionAnti-kt-R5Table},
but with $R = 0.4$.
\label{CrossSectionAnti-kt-R4Table} }
\end{table*}

In our study, we consider the inclusive process $p p \rightarrow$
\Wjjjj{} jets at an LHC center-of-mass energy of $\sqrt{s} = 7$ TeV.  We 
impose the following cuts: $E_T^{e} > 20$ GeV, $|\eta^e| < 2.5$,
$\ETsl > 20$ GeV, $\pt^\jet > 25$ GeV, $|\eta^\jet|<3$, 
and $M_T^W > 20$ GeV.  Here,
$p_T$ are transverse momenta; $\eta$,  pseudorapidities; and
$M_T^W$, the transverse mass of the
$e \nu$ pair. The missing transverse energy, $\ETsl$, corresponds to
the neutrino transverse energy, $E_T^\nu$.  Jets are defined using
the anti-$k_T$ algorithm~\cite{antikT} with parameter
$R = 0.5$, 
and are ordered in $\pt$.   (We also quote results for $R=0.4$.)
We use the CTEQ6M~\cite{CTEQ6M} parton
distribution functions and $\alpha_s$ at NLO, and the CTEQ6L1 set at LO.
Electroweak couplings are as in ref.~\cite{W3jDistributions}.

In \tab{CrossSectionAnti-kt-R5Table}, we present LO and NLO
parton-level cross sections for inclusive $W^-$-boson
production accompanied by zero through four jets.  We include
all subprocesses, using the leading-color virtual approximation 
only in \Wjjjj-jet production.  The upward scale-variation figures for
the NLO cross sections are quite small for \Wjjj- and \Wjjjj-jet
production, because the values at the central scale choice are close
to the maximum values across scale variations.  We also display the
ratios of the $W^+$ to $W^-$ cross sections, and the
``jet-production'' ratios of \Wmjn-jet to \Wmjnm-jet production.
Both kinds of ratios should be less sensitive to experimental and
theoretical systematics than the absolute cross sections.

The $W^+/W^-$ ratios are greater than one because the LHC is a $pp$
machine, and because the parton luminosity ratio $u(x)/d(x)$ exceeds
one. As the number of jets increases, production of a $W$ requires a
larger value of $x$, driving $u(x)/d(x)$ and hence the $W^+/W^-$
ratio to larger values.  These ratios have been discussed
recently~\cite{KomStirling} as a probe of certain new-physics
processes; our results extend the NLO analysis to $W$ production
accompanied by four jets.  This ratio changes very little under
correlated variations of the scale in numerator and denominator; hence
we do not exhibit such scale variation here. 

Standard lore~\cite{BerendsRatio}
says that the jet-production ratio should be roughly
independent of the number of jets.  The results for the
ratios displayed here for
$n>1$ are indeed consistent with this lore.  However,
they are rather sensitive to the experimental cuts, and
can depend strongly on $n$ when binned in the vector-boson
$p_T$~\cite{TeVZ}.
The \Wj-jet/\Wz-jet ratio is much smaller because of the
restricted kinematics of the leading contribution to \Wz-jet production.

In \tab{CrossSectionAnti-kt-R4Table}, we give
cross sections for narrower jets, using the anti-$k_T$ jet algorithm
with $R=0.4$.  For two or more jets, the LO cross sections are larger
than for $R=0.5$, and the effect increases with the number of jets. 
However, at NLO, the effect is greatly diminished; only for four jets
is the NLO cross section for $R=0.4$ significantly above that for
$R=0.5$.
The NLO jet-production ratio is somewhat larger for $R=0.4$, for $n>2$;
in contrast, the ratios of
$W^+$ to $W^-$ cross sections are unchanged within errors.

In \fig{W4ptFigure}, we show the $p_T$ distributions of the 
leading four jets in \Wmjjjj-jet production
at LO and NLO; the predictions are normalized to the central NLO prediction
in the lower panels.
With our central scale choice, there is a noticeable
shape difference between the LO and NLO distributions for the first
three leading jets, while the fourth-jet distribution is very similar at LO
and NLO.  Similarly, in \Wjjj-jet production, the $p_T$ distributions of
the leading two jets exhibit shape changes from LO to NLO, 
while the third-jet distribution does not~\cite{W3jDistributions}.

\ifpreprint
\begin{figure}[t]
\includegraphics[clip,scale=0.34]{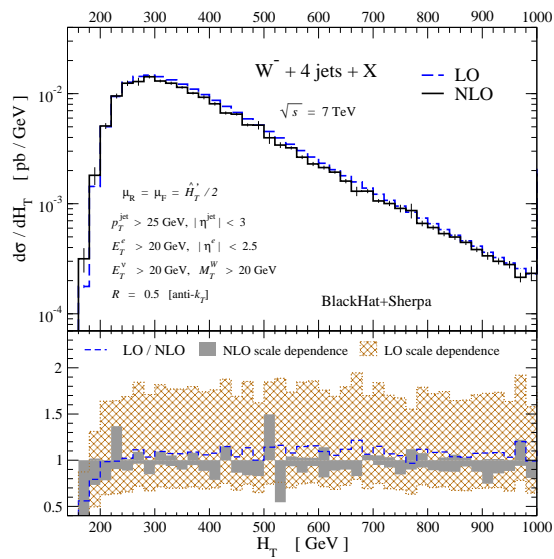}
\caption{The $H_T$ distribution for \Wmjjjj{} jets.
 }
\label{HTFigure}
\end{figure}

\Fig{HTFigure} shows the distribution of the total transverse
energy $H_T$, given by the scalar sum of the jet and lepton transverse
energies, $H_T = \sum_j E_{T,j}^{\rm jet} + E_T^e + E_T^\nu$.  We show
the NLO and LO predictions, along with their scale-dependence
bands. As in the $p_T$ distributions, the NLO band is narrower.
The shapes at LO and NLO are similar above 200 
GeV, where the integration errors are small.
\fi



The results of this study validate our understanding of the
\Wjjjj-jet process for typical Standard-Model cuts.  It will be
interesting, and necessary, to explore the size of corrections for
observables and cuts used in new-physics searches.

In order to compare our parton-level results to forthcoming
experimental data, the size of non-perturbative effects (such as
hadronization and the underlying event) needs to be estimated,
for example using LO parton-shower Monte Carlo programs.
As NLO parton-shower programs are
developed~\cite{POWHEGBOXMCNLO}, the virtual corrections computed here
should be incorporated into them.

A related process that contributes an irreducible background to
certain missing energy signals of new physics is \Zjjjj-jet
production.  We expect that the current \BlackHat{} along with
\SHERPA{} will allow us to compute NLO corrections to it, as well as
to other complex processes, thereby providing an unprecedented level
of theoretical precision for such backgrounds at the LHC.


\vskip .3 cm 

We thank Beate Heinemann, Johannes Henn and Rainer Wallny for helpful
discussions.  This research was supported by the US Department of
Energy under contracts DE--FG03--91ER40662, DE--AC02--76SF00515 and
DE--FC02--94ER40818.  DAK's research is supported by 
the European Research Council under
Advanced Investigator Grant ERC--AdG--228301.
HI's work is supported by a grant from the US
LHC Theory Initiative through NSF contract PHY--0705682.
 This research used
resources of Academic Technology Services at UCLA and of the National
Energy Research Scientific Computing Center, which is supported by the
Office of Science of the U.S. Department of Energy under Contract
No. DE--AC02--05CH11231.

\end{document}